# Streaming Balanced Graph Partitioning Algorithms for Random Graphs

Isabelle Stanton
University of California Berkeley
Berkeley, CA, USA
isabelle@eecs.berkeley.edu

## ABSTRACT

There has been a recent explosion in the size of stored data, partially due to advances in storage technology, and partially due to the growing popularity of cloud-computing and the vast quantities of data generated. This motivates the need for streaming algorithms that can compute approximate solutions without full random access to all of the data.

We model the problem of loading a graph onto a distributed cluster as computing an approximately balanced $k$-partitioning of a graph in a streaming fashion with only one pass over the data. We give lower bounds on this problem, showing that no algorithm can obtain an $o(n)$ approximation with a random or adversarial stream ordering. We analyze two variants of a randomized greedy algorithm, one that prefers the $\arg\max$ and one that is proportional, on random graphs with embedded balanced $k$-cuts and are able to theoretically bound the performance of each algorithms - the $\arg\max$ algorithm is able to recover the embedded $k$-cut, while, surprisingly, the proportional variant can not. This matches the experimental results in [25].

## 1. INTRODUCTION

Recent advances in storage technology and distributed computing have led to the phenomenon known as *big data*. There are several paradigm shifts involved in the big data movement. From a theoretical perspective, one is that traditional assumptions like full uniform random access to the data are no longer reasonable. This shift motivates the study of streaming algorithms.

One very natural type of 'big data' is graphs. There has been a lot of interest in distributed graph computation systems from many different communities - systems builders, database experts, and machine learning. This has led to the creation of a huge number of such systems including GraphLab [18], Pregel [20], Horton, Spark [26], Trinity, and the filtering technique for MapReduce to name but a few. In terms of distributing data, some of these systems support custom partitionings but the vast majority use a hashing method to produce a random cut as the default partitioning. From a systems perspective, this approach makes sense - it is fast and is easy to maintain. However, the network is far slower than local communication between processor cores. A random cut on a graph is a good approximation to the MAXCUT problem and is the exact opposite of what one should do if one cares about communication volumes. Even marginal improvements in the partitioning can lead to large improvements in run time for distributed algorithms [25].

The communication problem is a major motivator of the study of graph partitioning. The constraints of distributed computing and the fact that the graph data arrives as a stream means that traditional graph partitioning algorithms that assume full access to the data are no longer scalable solutions. In this paper, we consider the problem of finding an approximately balanced $k$-partitioning of a graph using a streaming algorithm with only one pass over the data as this models partitioning a graph while loading it onto a cluster.

Previous work addressed this problem from an experimental perspective. [25] evaluates 16 different partitioning heuristics on 21 different graphs to find how well each performs when compared with an offline partitioning heuristic (METIS [14]). A greedy algorithm assigns a vertex to the partition where it currently has the most edges. Surprisingly, a simple variant of greedy performed the best, even beating an adaptation of a local partitioning algorithm, EvoCut [3]. Also surprising was that adding randomization to the same algorithm caused it to perform significantly worse. Often, the addition of randomness often allows us to design more effective algorithms, not less. In this paper, we seek to provide a theoretical foundation for understanding these results and motivate further study into more sophisticated algorithms.

*Contributions.*

This paper focuses on developing a rigorous understanding of two greedy streaming balanced graph partitioning algorithms. We first give lower bounds on the approximation ratio that any streaming algorithm for balanced graph partitioning can obtain on both a random and adversarial ordering of the graph. In response to this lower bound, we focus our attention on a class of random graphs with embedded balanced $k$ cuts. We analyze our greedy algorithms by using a novel coupling to finite Polya Urn processes. This is very elucidating connection gives clear intuition as to why one algorithm performs well while the other does not. We finish with an experimental evaluation of the bounds attained by the theorems.

## 2. RELATED WORK

Many variants of graph partitioning have been studied since the earliest days of Computer Science. The variant considered in this paper, balanced $k$-cut, has been shown to be NP-hard by Andreev and Răcke [4], even when one relaxes the balance constraint. They also give an LP-based solution that obtains an $O(\log n)$ approximation. Another full-information solution was found by Even et al. who use an LP solution based on spreading metrics to also obtain an $O(\log n)$ approximation algorithm [12]. If one ignores the balance constraint, a popular approach is to use the top $k$ eigenvectors [22]. Recently, this approach was theoretically validated as an extension of Cheeger's inequality [16, 17]. One can also use any balanced 2-partitioning algorithm to obtain an approximation to a balanced $k$-partitioning when $k$ is a power of 2, losing at most an additional $\log n$ factor [5].

From a heuristic perspective, there are numerous full information graph partitioning systems available that do not have theoretical performance guarantees. These include METIS [14], PMRSB [7], and Chaco [13].

Another approach, relevant for our limited information setting, is *local partitioning* algorithms. The goal here is not to obtain a balanced cut but given a starting node to find a good cut around that node. Spielman and Teng were the first to develop this style of algorithm [24]. Anderson, Chung and Lang improved upon Spielman and Teng's work by using personalized PageRank vectors to find a good local cut [2]. Addressing the same problem, Anderson and Peres use the evolving graph process to obtain similar results [3]. While local partitioning is similar in spirit, it is not the same as a streaming algorithm.

The main focus of this paper is on streaming algorithms and there is significant related work in this area as well. First, noting the connection between graph partitioning and PageRank is Das Sarma et al.'s work on computing the PageRank of a graph with multiple passes [10]. Closer to our setting, Bahmani et al. incrementally compute an approximation of the PageRank vector with only one pass [6]. However, just computing the approximate PageRank vector is not sufficient for finding a graph partitioning with only one pass over the data. Das Sarma et al. extend their techniques to find sparse cut projections within subgraphs, again using multiple passes over the stream [9]. Cut projections are not the same as finding balanced cuts.

An alternate model, *semi-streaming*, assumes that we have $O(n\text{poly}\log n)$ storage space so that all vertices can be stored but the edges arrive in some order. In this setting, Ahn and Guha [1] give a one pass $\tilde{O}(n/\epsilon^2)$ space algorithm that sparsifies a graph such that each cut is approximated to within a $(1+\epsilon)$ factor. Kelner and Levin [15] produce a spectral sparsifier with $O(n \log n/\epsilon^2)$ edges in $\tilde{O}(m)$ time. While sparsifiers are a great way of reducing the size of the data, this reduction would then require an additional pass over the data to compute a partitioning which is out of the scope of the problem at hand. Finally, lower bounds are known with regards to the space complexity of both the problem of finding a minimum and maximum cut. Zelke [27] has shown that this cannot be computed in one pass with $o(n^2)$ space.

Finally, analyzing algorithms on random graph models has a long history. In particular, it is quite common to analyze graph partitionings on random graphs with planted partitions [21, 19]. This is done because recovering a planted partition is equivalent to finding the 'right' answer.

## 3. NOTATION AND DEFINITIONS

We now introduce the notation and definitions used throughout the rest of the paper. The *balanced graph partitioning problem* takes as input a graph $G$, an integer $k$ and an allowed imbalance parameter of $\epsilon$. The goal is to partition the vertices of $G$ into $k$ sets, each no larger than $(1+\epsilon)\frac{n}{k}$ vertices, while minimizing the number of edges cut.

### *Graph Models.*

A graph $G = (V, E)$ consists of $n = |V|$ vertices and $m = |E|$ edges. $\Gamma(v)$ is the set of vertices that a vertex $v$ neighbors. We consider graphs generated by two random models. The first, $G(n, p)$ is the traditional Erdös-Renyi model with $n$ vertices. The traditional definition is that each of the possible $\binom{n}{2}$ edges is included independently with probability $p$. At certain points in the proofs in Section 5, we modify this definition to make it better match our streaming model. In particular, we allow multiple edges in order to maintain independence in our analysis.

$G(\Psi, P)$ is a generalization of $G(n,p)$, due to McSherry [21], that allows the graph to have $l$ different Erdös-Renyi components, each with different parameters. Again, we have $n$ vertices. $\Psi : \{1, 2, \ldots n\} \to \{1, 2, \ldots l\}$ is a function mapping the vertices into $l$ disjoint clusters. Let $C_i$ refer to the set of vertices mapped to $i$, i.e. $\Psi^{-1}(i) = C_i$. $P$ is a $l \times l$ matrix where edges between vertices in $C_i$ are included independently with probability $P_{i,i}$ and edges between vertices in $C_i$ and $C_j$ are included with probability $P_{i,j}$. There are many ways for $G(\Psi, P)$ to generate graphs in $G(n,p)$ - $\Psi$ could map all vertices into the same cluster or we could have $P_{i,i} = P_{i,j} = p$ for all $i,j$. We make the same modification to the generative process as in $G(n,p)$ and allow multiple edges for clarity of the analysis.

*Probability Distributions.*
We only use variables drawn from a binomial distribution, where $X \sim B(n,p)$ is a random variable representing $n$ independent trials, each with probability $p$ of success.

## 3.1 Polya Urn Processes

The classical Polya Urn problem is: Given finitely many initial bins, each containing one ball, let additional balls arrive one at a time. For each new ball with probability $p$ create a new bin and put the ball in it. With probability $1-p$, place the ball in an existing bin with probability proportional to $m^\gamma$ where $m$ is the number of balls currently in that bin.

Many variants of the above process have been analyzed. In particular, Chung, Handjani, and Jungreis [8] analyze the *finite Polya urn process* where $p = 0$. The exponent $\gamma$ plays an important role in the behavior of this process. With $k$ bins, when $\gamma < 1$, in the limit, the load of each bin is uniformly distributed and each contains a $\frac{1}{k}$ fraction of the balls. When $\gamma > 1$, in the limit, the fractional load of one bin is 1. When $\gamma = 1$, the limit of the fractional loads exists but is distributed uniformly on the simplex.

Our proof technique will focus on connecting the streaming graph partitioning algorithms with the finite Polya urn process and use many of the results from [8]. We restate the results used here:

THEOREM 1 (THEOREM 2.1 FROM [8]). *Consider a finite Polya process with exponent $\gamma = 1$, $k$ bins and let $x_i^t$ denote the fraction of balls in the $i^{th}$ bin at time $t$. Then almost surely for each $i$, the limit $X_i = \lim_{t\to\infty} x_i^t$ exists. Furthermore these limits are distributed uniformly on the simplex $\{(X_1, X_2, \ldots X_k) : X_i > 0, X_1 + X_2 + \ldots + X_k = 1\}$.*

THEOREM 2 (THEOREM 2.2 FROM [8]). *Consider a finite $k$-bin Polya process with exponent $\gamma$ and let $x_i^t$ denote the fraction of the balls in bin $i$ at time $t$. Then a.s. the limit $X_i = \lim_{t\to\infty} x_i^t$ exists for each $i$. If $\gamma > 1$ then $X_i = 1$ for one bin, and $X_i = 0$ for all others. If $\gamma < 1$ then $X_i = \frac{1}{k}$ for all bins.*

LEMMA 1 (LEMMA 2.3 FROM [8]). *Given a finite or infinite Polya process with exponent $\gamma$ and an arbitrary initial configuration (i.e. finitely many balls arranged in finitely many bins), suppose we restrict attention to any particular subset of the bins and ignore any balls that are placed in the other bins. Then the process behaves exactly like a finite Polya process with exponent $\gamma$ on this subset of bins, though the process may terminate after finitely many balls.*

Lemma 1 is particularly important to our analysis as it forms the basis of an inductive argument to extend the analysis in [8] to $k$ bins from 2 bins. We also use the claim that a finite, arbitrary initial configuration does not affect the distribution in the limit.

## 3.2 The Streaming Model

We consider a streaming graph model where the vertices arrive in some order. The two stream orderings we consider are *adversarial* and *random*. For $n$ vertices, the set of permutations $S_n$ defines all possible orderings. For a random ordering, each permutation is picked with equal probability. An adversarial ordering is any probability distribution over the permutations, including one that picks the worst possible ordering for the algorithm.

When a vertex arrives so do all of its incident edges. Our goal is to generate a balanced vertex partitioning of the graph with $k$ partitions. The capacity of each partition, $C$, is enough to hold all the vertices, i.e. $kC = (1+\epsilon)n$. We assume an undirected graph since our evaluation metric, the number of edges cut, is not affected by the directionality of an edge.

We chose this model because we are concerned with the problem of loading data onto a cluster and partitioning at the same time. We assume that only one pass can be made over the data and the algorithm has access to the current load of each machine on the cluster and the location of each vertex that has been previously seen. A vertex is not moved after it has been placed into some partition.

# 4. LOWER BOUNDS

Given our streaming model, the first important question is whether any algorithm can do well on all graphs. The unfortunate answer is no. Intuitively, with only one pass, important edges may be hidden either intentionally by an adversary or unintentionally by randomness.

THEOREM 3. *One-pass streaming balanced graph partitioning with an adversarial stream order can not be approximated within $o(n)$.*

PROOF. Without loss of generality, we seek a balanced 2 partitioning. Consider a graph that is a cycle over $n$ vertices with edges such that $(i, i+1) \mod n \in E$ for $1 \le i \le n$. Let the ordering be all odd nodes, then all even, i.e. $1, 3, 5 \ldots n-1, 2, 4, 6 \ldots n$. Assume that $n$ is even. The optimal balanced partitioning cuts 2 edges. However, the given ordering reveals no edges until $\frac{n}{2}$ vertices arrive. Until the edges arrive, we have no way of distinguishing which vertices are 'near' each other. In

**Algorithm 1** arg max Greedy

**Input:** $G$, $k$, $C$, $\pi$

$P_1, \cdots, P_k = \emptyset$
**for** $t = 1, 2, \ldots n$ **do**
  **for** $i = 1, 2, \ldots k$ **do**
    $S_i = |\Gamma(\pi(t)) \cap P_i|$
    **if** $|P_i| = C$ **then**
      $S_i = 0$
  **if** all $S_i = 0$ **then**
    Pick $i$ from $\arg\min_{j\in[k]}\{|P_j|\}$ u.a.r.
  **else**
    Pick $i$ from $\arg\max_{j\in[k]}\{S_j\}$ u.a.r.
  $P_i = P_i \cup \pi(t)$

**Algorithm 2** Proportional Greedy

**Input:** $G$, $k$, $C$, $\pi$

$P_1, \cdots, P_k = \emptyset$
**for** $t = 1, 2, \ldots n$ **do**
  **for** $i = 1, 2, \ldots k$ **do**
    $S_i = |\Gamma(\pi(t)) \cap P_i|$
    **if** $|P_i| = C$ **then**
      $S_i = 0$
  **if** all $S_i = 0$ **then**
    Pick $i$ from $\arg\min_{j\in[k]}\{|P_j|\}$ u.a.r.
  **else**
    Pick $i$ proportional to $S_i$
  $P_i = P_i \cup \pi(t)$

particular, note that this ordering is indistinguishable from one where the odd vertices are given in a random order, or one where the odd nodes are interspersed with unconnected even nodes, i.e. $1, n-2, 3, n-4, 5, n-6 \ldots$. Thus, no algorithm can do better than cutting $\frac{n}{2}$ edges in expectation. This generalizes to $k$ partitions. □

THEOREM 4. *One-pass streaming balanced graph partitioning with a random stream order can not be approximated within $o(n)$.*

PROOF. Again, we seek a balanced 2 partition for a cycle graph with a random ordering. Consider the $t^{th}$ vertex to arrive in this ordering.

$$\Pr[t \text{ arrives with no edges }] = \Pr[\text{ both neighbors arrive after } t] = \frac{n-t}{n}\frac{n-t-1}{n-1}$$

so the number of vertices that we expect to arrive with no edges is

$$\mathbf{E}[\# \text{ with no edge}] = \sum_{t=1}^{n} \frac{t}{n}\frac{t+1}{n-1} \approx \frac{1}{n^2}\sum_{t=1}^{n} t^2 - t = \frac{1}{n^2}\left(\frac{n^3}{3} + \frac{n^2}{2} + \frac{n}{6} + \frac{n(n+1)}{2}\right)$$

Therefore, asymptotically, we expect $\frac{n}{3}$ vertices to arrive with no edges. As before, when a vertex arrives with no edges, we are not able to determine which other vertices it is 'near'. For each of these, we expect to cut 1 edge, providing us with our lower bound. □

In the following sections, we only analyze the algorithms for random orderings. In particular, we will show that for random graphs with higher degree and a planted partition, arg max Greedycan recover the partitioning.

# 5. ANALYSIS OF ALGORITHMS ON RANDOM GRAPHS

The experiments in [25] showed that one heuristic studied in the paper, Linear Deterministic Greedy (LDG), was clearly the best tried. However, another heuristic, Linear Randomized Greedy(LRG), differs only in that it selects a partition proportionally to the distribution of edges instead of from the maxima. In the experiments, LDG performed significantly better than LRG. This raises the question - can we theoretically explain the difference in performance? In this section, we will introduce slightly simpler variants, arg max Greedy (corresponding to LDG) and Proportional Greedy(corresponding to LRG), and analyze their performance on McSherry's random graph model. Our analysis will clearly demonstrate the difference observed in the experiments.

## 5.1 Algorithms

The two algorithms studied in this paper are very similar: when a vertex $v$ arrives, a score for each partition $P_i$ of the number of edges from $v$ to $P_i$, $S_i = |\Gamma(v) \cap P_i|$, is calculated. If the partition is full, its score is set to 0. If all scores are 0, then the vertex is assigned to some partition with minimal load. If a score is non-zero, then the arg max Greedy Algorithm assigns the vertex uniformly at random to a partition in $\arg\max S_i$. By contrast, the Proportional Greedy Algorithm uses the scores as a distribution, assigning the vertex to partition $i$ with probability $S_i / \sum S_j$.

The versions of these algorithms from [25] differ only in that the score for each partition is weighted by the current load of the partition, i.e. $S_i(1 - \frac{|P_i|}{C})$. In practice, the algorithms keep the partitions nearly balanced, meaning this tiebreaker is only used in cases of tied number of edges and when there are no edges where [25] prefers the least-loaded partitions.

One of the key insights of this paper is that when these algorithms are used on random graphs, we can write both down as random processes. In particular, we can let the random process generate the graph while also partitioning it at the same time. This reduction will be discussed in Section 5.3. The proof proceeds by analyzing the random process versions of the algorithms, rather than those given in Algorithms 1 and 2.

The random processes generate a multi-edge $G(n, p)$ graph. For the extended $G(\Psi, P)$ analysis, we will only

**Algorithm 3** arg max Greedy Process on $G(n,p)$

**Input:** $p$

Set $P_1, P_2, \dots P_k = \emptyset$
**for** $t = 1, 2, \dots n$ **do**
  For $1 \le i \le k$, draw $E_i^{(t)} \sim B(|P_i|, p)$
  **if** $\sum_{i=1}^{k} E_i^{(t)} = 0$ **then**
    Assign $t$ to $\arg\min_{j\in[k]}\{|P_j|\}$
  **else**
    Assign $t$ to $\arg\max_{j\in[k]}\{E_j^{(t)}\}$

**Algorithm 4** Proportional Greedy Process on $G(n,p)$

**Input:** $p$

Set $P_1, P_2, \dots P_k = \emptyset$
**for** $t = 1, 2, \dots n$ **do**
  For $1 \le i \le k$, draw $E_i^{(t)} \sim B(|P_i|, p)$
  **if** $\sum_{i=1}^{k} E_i^{(t)} = 0$ **then**
    Assign $t$ to $\arg\min_{j\in[k]}\{|P_j|\}$
  **else**
    Assign $t$ to $P_i$ with probability $E_i^{(t)} / \sum_{j=1}^{k} E_j^{(t)}$

consider Algorithm 1 and the correctly modified version of Algorithm 3. The modification is only with the generation of the $E_i^{(t)}$ and will be discussed later.

## 5.2 Result and Proof Outline

The rest of the paper will focus on proving the following two statements. The first is that the Proportional Greedy Algorithm can not recover an embedded partition in a $G(\Psi, P)$ graph, no matter what the parameters are or how big the graph is. By contrast, the second result is that the arg max Greedy Algorithm can recover the embedded partition, provided the components are dense enough, the cut between them is sparse enough, and there are enough components.

Theorem 5. *Let $p$ be the probability of edges within components and $q$ be the probability of edges between components. Given a $G(\Psi, P)$ graph with $l > k \log k$ equally sized components where $p > \frac{2\log n}{|C|}$, $p > 3(k + \sqrt{k}+1)lq$, and $q = O((k^{2.4}\log l)^{-1})$, arg max Greedy Algorithm will recover an embedded partition from a random stream ordering.*

The proof proceeds in several stages. First, we ignore the capacity constraint and consider Algorithms 3 and 4 on a single $G(n,p)$ component. Does the algorithm eventually learn it is a component and place it in the same partition? We show that Algorithm 4 is equivalent to a finite Polya urn process with $\gamma = 1$ and distributes the component over all the partitions. By contrast, Algorithm 3 can be coupled to a finite Polya urn process with $\gamma > 1$. It will asymptotically place the entire $G(n,p)$ component in one partition. This argument starts with 2 partitions and is extended to $k$ bins using an induction argument.

That Algorithm 3 will correctly (not) partition a connected component forms the basis of our argument that it can be extended to the $G(\Psi, P)$ model. Intuitively, with the correct parameters, each component of $G(\Psi, P)$ will be placed in a single partition. The primary technical difficulties faced are the inclusion of the capacity constraint, requiring bounds on the component sizes, and the addition of intra-cluster edges, which serve to 'confuse' the algorithm about to which component a vertex belongs. By setting the parameters of the model correctly, we can overcome these challenges.

## 5.3 Analysis on a Single $G(n,p)$ Component

We now analyze Algorithms 3 and 4. These are obtained from Algorithms 1 and 2 by considering the process in terms of Polya urns. As a reminder, the finite Polya urn process has $k$ bins and the $t^{th}$ ball is assigned to bin $i$ with probability proportional to $(m_i^{(t)})^\gamma$ where $m_i^{(t)}$ is the load of the $i^{th}$ bin at time $t$.

Translating Algorithm 1 and 2 to Polya Urn processes involves identifying each ball with a vertex and each bin with a partition. There are two primary differences from the standard Polya Urn process. First, with probability $(1-p)^t$, the $t^{th}$ vertex (ball) does not have edges to vertices already seen and it is placed in the least loaded partition (urn). The second is that we do not assign the vertex (ball) based on the load of the partition (urn) but instead on a binomial random variable based on the load. Specifically, let $E_1^{(t)}, \dots E_k^{(t)}$ be the random variables representing the number of edges to each of the $k$ partitions. Each $E_i^{(t)}$ is drawn from $B(m_i^{(t)}, p)$. The following connection is how we created Algorithms 3 and 4.

- Algorithm 1 assigns the vertex to a partition in $\arg\max_{j\in[k]}\{E_j^{(t)}\}$, breaking ties at random.
- Algorithm 2 assigns it to bin $i$ proportional to $E_i^{(t)}$

*Algorithm 4 Analysis.* Consider the total number of edges from vertex $t$ as a random variable $E^{(t)} \sim B(t,p)$. Each edge is distributed according to $m_i^{(t)}$ i.e. with probability $\frac{m_i^{(t)}}{t}$ it connects to the $i^{th}$ bin. Each of the $E^{(t)}$ edges are distributed i.i.d. and are given equal weight so Algorithm 2 assigns balls proportional to $(m_i^{(t)})^\gamma$ where $\gamma = 1$.

Theorem 6. ***Algorithm 4 on*** $G(n,p)$ *Let $0 \le p < 1$. Let $x_i^t$ be the fractional load of partition $i$ at time $t$ of Algorithm 4. Then almost surely $\lim_{t\to\infty} x_i^t = X_i$ exists and for all $i$, $X_i > 0$.*

PROOF. We show that when there are edges, this process is exactly a finite Polya urn process with $\gamma = 1$. The result then follows directly from Theorem 1. Let there be $k$ bins. At time $t$, each has load $m_i^{(t)}$. Let $E^{(t)}$ be the total number of edges drawn by the process. Assume $E^{(t)} > 0$ as $E_t^{()} = 0$ will be dealt with later. Recall that we allow multiple edges in our model, so consider the edges being distributed to the $k$ partitions with replacement, i.e. each of the $E^{(t)}$ edges goes to partition $i$ with probability $\frac{m_i^{(t)}}{t-1}$. Let $E_i^{(t)}$ be the number to partition $i$. Note that $\sum_{i=1}^{k} E_i^{(t)} = E^{(t)}$. Now $\Pr[\text{Algorithm 4 picks bin } i] = E_i^{(t)}/E^{(t)}$. However, $E_i^{(t)} \sim B(E^{(t)}, \frac{m_i^{(t)}}{t-1})$, showing that this assignment is proportional to $m_i^{(t)}$ as desired. This is exactly a finite Polya urn process with $\gamma = 1$.

The remaining details concern the modification of the process when $E^{(t)} = 0$. In this case, the algorithm will assign the vertex to the least loaded bin. If this situation has a constant probability throughout the process, then it is making the distribution of the balls more uniform, and satisfy the theorem statement that all bins contain a non-zero fraction of the balls. If it is the case that this becomes unlikely as the process progresses, i.e. $p > \frac{\log n}{n}$, then we can apply Theorem 1 and Lemma 1 from [8] to say that after $O(\frac{\log n}{p})$ vertices have arrived, we begin the $\gamma = 1$ Polya Urn process with an arbitrary finite initial configuration. From Theorem 1, we get that $X_i > 0$ for all $i$. □

We conclude that the randomized algorithm does not have a concentration result. No matter the value of $p$ or the size of the graph, for a $G(n,p)$ component, the Proportional Greedy algorithm will not learn that it is a component and instead distributes it over all partitions.

COROLLARY 1. *Given a single isolated $G(n,p)$ component, for any value $p$, Algorithm 2 will distribute this component over all $k$ partitions.*

*Algorithm 3 Analysis.* The key insight about why Algorithm 3 provides a concentration result is that by preferring the arg max of the distribution of edges, once some partition has a slightly higher load than the other it is very likely to be assigned the next vertex. As the gap in the loads grow, the larger partition becomes increasingly more likely to receive the next vertex until it is impossible for the smaller partition to compete. However, there are a few challenges.

The first is that with probability $(1-p)^t$ the $t + 1^{th}$ vertex will not have any edges to previously seen vertices. In this case, it is automatically placed in the least loaded bin. When this happens, it decreases the gap in the loads. If it happens too often, the gap will not grow. Since $(1-p)^t \approx e^{-pt}$, once $t = O(\frac{\log n}{p})$, this does not happen with high probability, provided $p > \frac{\log n}{n}$. We only expect $\frac{1}{p}$ vertices to arrive with no edges and they are concentrated at the beginning of the process when $t < \frac{1}{p}$.

The second challenge is that when the vertex has 1 edge, the arg max distribution is the same as Algorithm 4. However, this can be dealt with in the same manner as having no edges. Again, we expect $\frac{1}{p}$ vertices to have only 1 edge and primarily when $\frac{1}{p} \leq t \leq \frac{2}{p}$. Therefore, we need $p > \frac{2\log n}{n}$.

The final challenge is that we are not be able to couple Algorithm 3 to a finite Polya urn process with $\gamma > 1$ until $\frac{2}{p}$ vertices have arrived, meaning we do not start with a uniform load distribution. Lemma 1 shows that we can start with an arbitrary finite initial configuration and obtain the same concentration results.

THEOREM 7. *Let $p$ be any value between $\frac{2\log n}{n}$ and 1. Let $x_i^t$ be the fractional load of partition $i$ at time $t$ of Algorithm 3. Then almost surely $\lim_{t\to\infty} x_i^t = X_i$ exists and one $X_j = 1$, while all others are 0.*

This statement follows from Theorem 2. Our analysis for Algorithm 3 relies on the probability that bin $i$ will receive a ball at time $t$ or

$$\Pr\left[E_i^{(t)} = \arg\max_{j\in[k]}\{E_j^{(t)}\}\right]$$

for $E_i^{(t)} \sim B(m_i^{(t)}, p)$. It is intuitive that bins with a higher load should have a much higher probability of being the arg max, yet the binomial distribution does not have a nice closed form expression for $\Pr[X \geq k]$. Even if we condition on $E^{(t)} = \sum_{i=1}^{k} E_i^{(t)} = x$ so we can express the $E_i^{(t)}$ as a multinomial distribution, a nice closed form solution eludes us.

Therefore, our proof consists of several lemmas.

LEMMA 2. *Given a $G(n,p)$ graph with $p > \frac{2\log n}{n}$, after $O(\frac{\log n}{p})$ steps, Algorithm 3 with 2 partitions can be coupled to a finite Polya urn process with $\gamma > 1$.*

PROOF. Let $A = E_1^{(t)}$ and $B = E_2^{(t)}$ and $A^j, B^j$ be the loads conditioned on the fact that $E^{(t)} = j$ i.e. $A^j + B^j = j$. Let $\delta$ be the comparative advantage of $A$ over $B$, i.e. $\frac{1}{2} + \delta = \frac{m_1^{(t)}}{t}$ and $\frac{1}{2} - \delta = \frac{m_2^{(t)}}{t}$. We want to analyze $\Pr\left[A^j > B^j\right]$.

$$\Pr\left[A^j > B^j\right] = \sum_{i=\lfloor j/2\rfloor+1}^{j} \binom{j}{i} (\frac{1}{2}+\delta)^i (\frac{1}{2}-\delta)^{j-i}$$

$$= (\frac{1}{2}+\delta)^{\lfloor j/2\rfloor+1} \sum_{i=\lfloor j/2\rfloor+1}^{j} \binom{j}{i} (\frac{1}{2}+\delta)^{i-\lfloor j/2\rfloor-1} (\frac{1}{2}-\delta)^{j-i}$$

$$= (\frac{1}{2}+\delta)^{\lfloor j/2 \rfloor+1} \sum_{i=0}^{\lfloor j/2 \rfloor} \binom{j}{i - \lfloor j/2 \rfloor} (\frac{1}{2}+\delta)^{i} (\frac{1}{2}-\delta)^{j-\lfloor j/2 \rfloor -1-i}$$

$$= (\frac{1}{2}+\delta)^{\lfloor j/2 \rfloor+1} \sum_{i=0}^{\lfloor j/2 \rfloor} \binom{j}{i} (\frac{1}{2}+\delta)^{\lfloor j/2 \rfloor - i} (\frac{1}{2}-\delta)^{i}$$

We similarly express $\Pr\left[B^j > A^j\right]$ as follows.

$$\Pr\left[B^j > A^j\right] = \sum_{i=\lfloor j/2 \rfloor+1}^{j} \binom{j}{i} (\frac{1}{2}-\delta)^{i} (\frac{1}{2}+\delta)^{j-i}$$

$$= (\frac{1}{2}-\delta)^{\lfloor j/2 \rfloor+1} \sum_{i=0}^{\lfloor j/2 \rfloor} \binom{j}{i} (\frac{1}{2}-\delta)^{\lfloor j/2 \rfloor - i} (\frac{1}{2}+\delta)^{i}$$

Because $\frac{1}{2}+\delta > \frac{1}{2}-\delta$, we have that $\sum_{i=0}^{\lfloor j/2 \rfloor} \binom{j}{i} (\frac{1}{2}+\delta)^{\lfloor j/2 \rfloor - i} (\frac{1}{2}-\delta)^{i} > \sum_{i=0}^{\lfloor j/2 \rfloor} \binom{j}{i} (\frac{1}{2}-\delta)^{\lfloor j/2 \rfloor - i} (\frac{1}{2}+\delta)^{i}$. Therefore,

$$\Pr\left[A^j > B^j\right] > \frac{(\frac{1}{2}+\delta)^{\lfloor j/2 \rfloor+1}}{(\frac{1}{2}-\delta)^{\lfloor j/2 \rfloor+1}} \Pr\left[B^j > A^j\right].$$

From this, and the fact that these two quantities sum to 1, we conclude that

$$\Pr\left[A^j > B^j\right] > \frac{(\frac{1}{2}+\delta)^{\lfloor j/2 \rfloor+1}}{(\frac{1}{2}+\delta)^{\lfloor j/2 \rfloor+1} + (\frac{1}{2}-\delta)^{\lfloor j/2 \rfloor+1}}$$

This lower bound is the probability that the ball goes in urn 1 in a Polya process with $\gamma = \lfloor j/2 \rfloor + 1$. When $j \geq 2$, we can couple our process to a finite Polya urn process with a desirable concentration result. We remove the conditioning on $E^{(t)} = j$ to get $\Pr[A > B]$

$$\Pr[A > B] = \sum_{j=1}^{t} \binom{t}{j} p^j (1-p)^{t-j} \Pr\left[A^j > B^j\right] \quad (1)$$

The only case where we are mixing in a process that has an undesirable exponent ($\gamma = 1$) is when $j = 0$ or 1. The probability of this case is less than $\frac{1}{n}$ when $t > \frac{2\log n}{p}$. According to Lemma 1, this constitutes a finite arbitrary configuration and the concentration results hold after $t > \frac{2\log n}{p}$. □

The above proof shows that, at some point, the algorithm can be coupled with a finite Polya urn process with $\gamma > 1$. However, we need Lemma 1 from [8] to show that the initial configuration when the process takes off does not affect the concentration results. Moreover, we bound the total expected number of vertices to arrive with $j = 0$ or 1 by

$$\frac{1-(1-p)^n + 1 - p}{p} \approx \frac{2 - e^{-pn} - p}{p} \leq \frac{2}{p}.$$

Combining Lemma 1 and 2 shows that for 2 partitions Algorithm 3 will concentrate the process into 1 bin. In order to extend the process to $k$ partitions, we present the following Lemma. It follows the proof technique of Theorem 2 in [8] and utilizes Lemma 1

LEMMA 3. *Consider Algorithm 3 with k partitions on a $G(n,p)$ graph with $p > \frac{2\log n}{n}$. Let $x_i^t$ be the fractional load of the $i^{th}$ partition at time t. Then a.s. the limit $X_i = \lim_{n,t\to\infty} x_i^t$ exists for each i. For exactly one i, $X_i = 1$.*

PROOF. To extend the analysis of Lemma 2 from 2 partitions to $k$, we use induction and condition on each pair of bins. Of the $k$ bins, select 2 and call them $A$ and $B$. We modify Lemma 2's Equation 1 by substituting

$$\Pr\left[E^{(t)} = j\right] = \binom{t}{j} p^j (1-p)^j$$

with

$$\Pr\left[E^{(t)} = j | A \text{ or } B \text{ is in the argmax}\right].$$

Given that our coupling to the Polya Urn process is unaffected, we just must show that

$$\Pr\left[E^{(t)} = 0, 1\right] > \Pr\left[E^{(t)} = 0, 1 | A \text{ or } B \text{ is in the argmax}\right].$$

The $E^{(t)} = 0$ case is simple since

$$\Pr\left[E^{(t)} = 0 | A \text{ or } B \text{ is the max}\right] = 0$$

since we only use the argmax process when $E^{(t)} \geq 1$ (otherwise we would have assigned the vertex to the least loaded partition). When $E^{(t)} = 1$, this is equivalent to exactly 1 edge being placed and the probability that, of the $k$ bins, it selects an endpoint in $A$ or $B$ is exactly $\frac{m_A^{(t)} + m_B^{(t)}}{t}$. Thus

$$\Pr\left[E^{(t)} = 1 | A \text{ or } B \text{ is the max}\right] =$$

$$\frac{m_A^{(t)} + m_B^{(t)}}{t} \binom{t}{1} p (1-p)^{t-1} \leq$$

$$\binom{t}{1} p (1-p)^{t-1} = \Pr\left[E^{(t)} = 1\right]$$

The result now follows from Theorem 2. □

*Proof of Theorem 7:* Combining Lemmas 2, 1, and 3, we conclude that Algorithm 3, with $k$ partitions, will asymptotically approach a fractional load of 1 in one partition when run with $p > \frac{2\log n}{n}$. □

COROLLARY 2. *Given a single $G(n,p)$ component, for any value $p > \frac{2\log n}{n}$, Algorithm 1 will eventually concentrate this component into 1 partition as $n \to \infty$.*

This analysis leaves open the question of how long the process must run before one partition dominates the others. This question has been studied by Drinea, Frieze and Mitzenmacher [11]. While they analyze the convergence rates for 2 bins, the proofs can be extended to $k$ bins via the union bound. In the theorem $B_0$ is the name for one of the two bins and all-but-$\delta$ dominant means that $B_0$ contains at least a $1-\delta$ fraction of the balls thrown. $\epsilon_0$ is the initial amount that the two bins are separated by after $n$ balls and is a constant depending on $\lambda$, say $\frac{1}{100\lambda}$.

THEOREM 8 (THEOREM 2.4 FROM [11]). *Assume that we throw balls into the system until $B_0$ is all-but-$\delta$ dominant for some $\delta > 0$. Then, if $\lambda > 1$, with probability $1 - e^{\Omega(n_0)}$, $B_0$ is all-but-$\delta$ dominant when the system has $2^{x+z}n_0$ balls, where $x = \log_{1+\frac{\lambda-1}{5+4(\lambda-1)}} \frac{0.4}{\epsilon_0}$ and $z = \log_{\frac{2\lambda}{\lambda+1}} \frac{0.1}{\delta}$.*

Lemma 4 extends this theorem to $k$ bins.

LEMMA 4 (LEMMA 4.1 FROM [11]). *Suppose that when $n$ balls are thrown into a pair of bins, the probability that neither is all-but-$\delta$ dominant is upper-bounded by $p(n,\delta)$. Here, we assume $p(n,\delta)$ is non-increasing in $n$. Then when $1+kn/2$ balls are thrown into $k$ bins, the probability that none is all-but-$\gamma$ dominant is at most $\binom{k}{2}p(n,\delta)$ for $\gamma = \delta/(\delta + (1-\delta)/(k-1))$*

To summarize these results on the convergence rate, we find that the attachment process starts in earnest after $\frac{1}{p}$ vertices have arrived. After $\frac{2}{p}$ vertices have arrived, we claim the exponent in the process is greater than 1. From Lemma 4 the probability we do not get an all-but-$\epsilon$ domination is inversely polynomial in the number of partitions, $1/\epsilon$ and the number of vertices. The bound given by Theorem 8 holds for $\lambda = 2$ but is loose since $\lambda$ value increases every after every round of $\frac{1}{p}$ vertices.

*Comparisons.* From these results, we conclude that the reason that Algorithm 2 fails to concentrate the component is the strict proportionality of its assignments. If instead it used any exponent greater than 1 on its scores, i.e. assign to $i$ proportional to $S_i^\gamma$, the concentration result would hold. In particular, there is a huge spectrum of greedy algorithms of the style of arg max Greedy and Proportional Greedy. Amongst these, arg max Greedy provides the strongest possible preference towards concentration.

## 5.4 Extending to $G(\Psi, P)$ graphs and capacity constraints

We showed that with no capacity constraints the arg max Greedy approach is able to asymptotically place a single $G(n,p)$ component into one partition. Specifically, while it will initially place vertices in all partitions, once we begin to see edges, the algorithm concentrates the component into one partition. By contrasts, the Proportional Greedy approach always cuts the component into $k$ pieces. We would like to extend this analysis for arg max Greedy to graphs that consist of many good clusters but face two challenges - the capacity constraints and the 'bad' inter-cluster edges.

These two challenges motivate our restrictions to both $\Psi$ and $P$. The capacity constraint can be violated if clusters are of size $c$ and the capacity is $\mathcal{C}$ and more than $\frac{\mathcal{C}}{c}$ communities chose a specific bin to form their large component. From the traditional analysis of throwing $m$ balls into $n$ bins, we know that the expected maximum load (with high probability) is $\frac{\log n}{\log\log n}$ when $m = n$ and $O(\frac{m}{n})$ when $m > n \log n$ [23]. If we can argue that for each cluster the location of its large component is chosen uniformly at random from the bins, then we can use the balls and bins maximum load analysis to argue that if each cluster is small enough, the slack required, $C = (1+\epsilon)\frac{n}{k}$, is also small. We also require a small amount of slack in the capacities to account for initial mistakes. These mistakes are the result of not seeing edges at the beginning of the process.

For simplicity, our proof will proceed by first assuming that all of the clusters, $C_i$, are of the same size and that $q$, the probability of inter-cluster edges, is 0. This will allow us to deal with running $l$ finite Polya Urn processes simultaneously and independently. After this, we show a non-zero bound on $q$ that will bound the probability of the process failing to find a cut on the inter-cluster edges small. Finally, the assumption that the $C_i$ are of equal size can be relaxed by adjusting the parameters in $P$ appropriately.

LEMMA 5. *Given a $G(\Psi, P)$ graph where $P_{i,j} = 0$, and $\forall i, |P_{i,i}| > 2\log n/|C_i|$, let $x_j^{(i)(t)}$ be the fraction of $C_i$ that partition $j$ holds at time $t$. With no capacity constraints, Theorem 7 will guarantee that, as $n$ grows, for each cluster $i$, if $\lim_{t\to\infty} x_j^{(i)(t)} = X_j^{(i)}$, then for some $j$, $X_j^{(i)} = 1$ while all others are 0.*

PROOF. This follows directly from Theorem 7 and the fact that when $P_{i,j} = 0$, the individual components can not interact with one another. □

Next, we relax the constraint that there are no edges between components to obtain a bound that still does not necessarily respect capacity constraints.

LEMMA 6. *Given a $G(\Psi, P)$ graph with $P_{i,i} = p$, $P_{i,j} = q$, and all $l$ clusters of equal size $|C_i| = |C_j|$ and $p > \frac{2\log n}{|C_i|}$. Let $x_j^{(i)(t)}$ be the fraction of $C_i$ that partition $j$ holds at time $t$. With no capacity constraints and $k$ partitions, if $p > 3(k + \sqrt{k} + 1)lq$ then for each cluster $i$, if $\lim_{t\to\infty} x_j^{(i)(t)} = X_j^{(i)}$, then for some $j$, $X_j^{(i)} = 1$ while all others are 0.*

Our goal is to bound the number of 'bad' inter-cluster edges away from the number of 'good' intra-cluster edges.

We assume worst case distributions so these bounds can safely be relaxed in practice.

Consider component $C_i$. A natural condition is that there are more expected intra-cluster edges than inter-cluster so $p|C_i| > q(n - |C_i|)$. We require a few more properties. The first is that the inequality holds with reasonable probability so $p|C_i| - \sqrt{p|C_i|} > q(n-|C_i|) + \sqrt{q(n - |C_i|)}$. The second is that we maintain the separation at every step of the execution of the process so $p|C_i|\frac{t}{n} - \sqrt{p|C_i|\frac{t}{n}} > q(n-|C_i|)\frac{t}{n} + \sqrt{q(n-|C_i|)\frac{t}{n}}$. Finally, we also need that the total number of bad edges should be no more than the arg max of the good edges as this guarantees that the bad edges will not affect the concentration results for each component. This adds a factor of $k$ to the bound so we must always guarantee there are at least $k$ 'good' edges for each 'bad' edge.

*Proof of Lemma 6:* Let the edges from a vertex to its own component be 'good' edges and its external edges be 'bad'. The separation between the good edges and bad edges can be achieved through the use of Chernoff bounds. In particular, at time $t$, we expect that $|C_i|\frac{t}{n}$ vertices in $C_i$ will have arrived already. Using a Chernoff bound to justify using the expectation, we claim that with probability at least $1-\delta$. Let the next vertex, $v$, be from $C_i$ . Let $E_i^{(t)}$ be the total number of edges from $v$ to the $C_i$ vertices that have already arrived.

$$E_i^{(t)} > p|C_i|\frac{t}{n} - \sqrt{\log(1/\delta)p|C_i|\frac{t}{n}}.$$

The bad edges, $B^t$, are drawn from $B(q, (n-|C_i|)\frac{t}{n})$. For clarity, we approximate $n - |C_i|$ as $l|C_i|$. Again, with probability at least $1 - \delta$, we claim that

$$B^t < ql|C_i|\frac{t}{n} + \sqrt{\log(1/\delta)ql|C_i|\frac{t}{n}}.$$

We set $\delta = 1/e$ to obtain constant probability at least 1/2. This assumption is supported by the experimental results in the next Section. We include bounds that hold with high probability in the Appendix.

To add the constraint that the bad edges are less than the $\arg\max\{E_i^{(t)}(j)\}$, we note that the worst case is that all of the bad edges connect to one partition. This can happen if the rest of the graph may not be evenly distributed over the partitions, or we are observing a deviation in the distribution of bad edges. Given this it is sufficient that the number of bad edges is bounded away from the average number of good edges, so we use the condition that

$$p|C_i|\frac{t}{n} - \sqrt{p|C_i|\frac{t}{n}} > k[ql|C_i|\frac{t}{n} + \sqrt{ql|C_i|\frac{t}{n}}]$$

To extract meaningful restrictions on $p$ and $q$ from this equation, we note that $p|C_i|\frac{t}{n} - \sqrt{p|C_i|\frac{t}{n}} > k$ when $t > \frac{(k+\sqrt{k}+1)n}{p|C_i|}$. Similarly, $ql|C_i|\frac{t}{n} + \sqrt{ql|C_i|\frac{t}{n}} < 1$ when $t < \frac{(1/2(3-\sqrt{5})n}{ql|C_i|}$. We find that $\frac{(k+\sqrt{k}+1)n}{p|C_i|} < \frac{(1/2(3-\sqrt{5})n}{ql|C_i|}$ exactly when $p > (k+\sqrt{k}+1)lq/(\frac{1}{2}(3-\sqrt{5}))$. Simplifying, $p > 3(k+\sqrt{(k)}+1)lq$ is sufficient. The gap between the left and right hand sides is monotonically increasing after this point, guaranteeing that all decisions will be made correctly with constant probability. □

Provided $k > 2$, $k+\sqrt{k}+1 < 2k$ so this bound is more simply $p > 3*2klq = 6klq$. If we make stronger assumptions about the distribution of the vertices within the bins at any finite time, i.e. that they are approximately balanced, then we can drop the $(k+\sqrt{k}+1)$ factor and obtain that $p > 3lq$ is sufficient.

The remaining technical point is the capacity constraints. Given that no aspect of the algorithm is dedicated towards load balancing when edges exist, our only hope can be that the components concentration points are distributed uniformly over the partitions. If this is the case then a standard balls-and-bins analysis will tell us how many components are assigned to each partition. In particular, if $n$ balls are thrown into $n$ bins, we expect the max load to be $\log n$ balls. However, if $n \log n$ balls are thrown into $n$ bins, we expect the max load to be $O(\log n)$. With more than $n \log n$ balls, the maximum load approaches the average.

This approach requires that we be able to argue that the inter-component edges have no affect on the concentration location for each component. This is clear when $q = 0$ and there are no 'bad' edges, the location of the concentration of each component is uniform because of the random ordering of the stream. Similarly, if $p = 1$ then the component is located exactly where the first vertex in the component is placed.

For other values of $p$ and $q$ we must use a more sophisticated argument. In particular, we can exploit the gap between $p$ and $q$ to argue that many intra-component edges are seen before any inter-component edges. If the process has run long enough that we can use Lemma 4 to argue that for each component, one partition contains a bit more than half of the vertices that have arrived, then we can argue that the arg max is never changed by the presence of 'bad' edges and that the processes do not affect each other.

LEMMA 7. *Given a $G(\Psi, P)$ graph with $P_{i,i} = p$, $P_{i,j} = q$ satisfying both Lemma 6 and $q = O((k^{2.4}\log l)^{-1})$, with the number of clusters $l > k \log k$ and all clusters of equal size $|C_i|$, with high probability the maximum load of the partitions is bounded by $(1+\epsilon)\frac{n}{k}$, where $\epsilon$ is a function of $p$, $l$ and $k$.*

PROOF. We first establish that the locations of the concentration for each component is uniformly distributed. This can be done by arguing that the partition that contains the maximum for each component is all-but-$\delta$ dominant and applying Theorem 8 and Lemma 4 to obtain that $q = O((k^{2.4} \log l)^{-1})$. The exact calculation is included in the Appendix.

Given the components are uniformly distributed over the partitions, this is a 'balls-and-bins' process with $l$ balls and $k$ bins. If $l = ck \log k$ then with high probability the maximum load is $d_c \log k$ where $d_c$ is a constant depending on $c$ [23]. When $l >> k \log k$ with high probability the maximum load is at most $\frac{l}{k} + \sqrt{2\frac{l}{k}\log k}$. From these results, we conclude that the clusters will be nearly evenly distributed amongst the bins.

Finally, $\epsilon$ needs to be set so that the capacity constraints will not be violated by either of the two sources. The first is the distribution of vertices before any edges appear. This is in expectation $\frac{1}{p}$ vertices, and each partition will hold $\frac{1}{pk}$ of them. The other source of slack required is the exact maximum load. This constant depends on $l$'s relationship to $k$ and can be obtained from [23]. □

The number of vertices required for the above argument to always hold is quite high in the analysis.

# 6. EXPERIMENTAL EVALUATION

The proofs in the previous sections show that for a certain range of parameters and size graph, the algorithm succeeds in recovering a good partitioning. It leaves open some interesting questions that can be experimentally evaluated:

- What is the relationship between $\epsilon$, the load balancing factor in Lemma 7 and $k$, the number of partitions, and $l$, the number of components?
- How tight are the bounds? It is necessary that the density of edges within components is $p > \frac{2 \log n}{|C_i|}$ or that the gap between $p$ and $q$, the probability of edges between components is at least $p > 6klq$?
- Are the convergence rates tight? For what size graph do we begin to recover the partitioning?
- When we are asymptotically recovering the partitioning, can we quantify how many mistakes we are making, i.e. how many vertices are separated from their components at the end of the process?

These questions are ideals candidate for experimental simulation. In fact, experimental results here can lead to a much better understanding of the algorithm than theoretical worst case bounds. In the following, using values that satisfy Lemma 6, we generate $G(\Psi, P)$ graphs and see how well arg max Greedy recovers the embedded cut.

*Evaluation.* Given a setting of the parameters, we generate a random $G(\Psi, P)$ graph and run the algorithm 25 times, each with a different random ordering. After each run, for each component in the $G(\Psi, P)$ graph, we its largest part in the partitioning i.e. if $C_i$ is the component, and $P_1, P_2, \cdots P_k$ the final partitioning, we calculate $\max_{j \in k} |C_i \cap P_j|/|C_i|$. The theorems predicts that for all components, this value approaches 1 as the graph grows. Note that it can never be worse than $\frac{1}{k}$ for $k$ partitions.

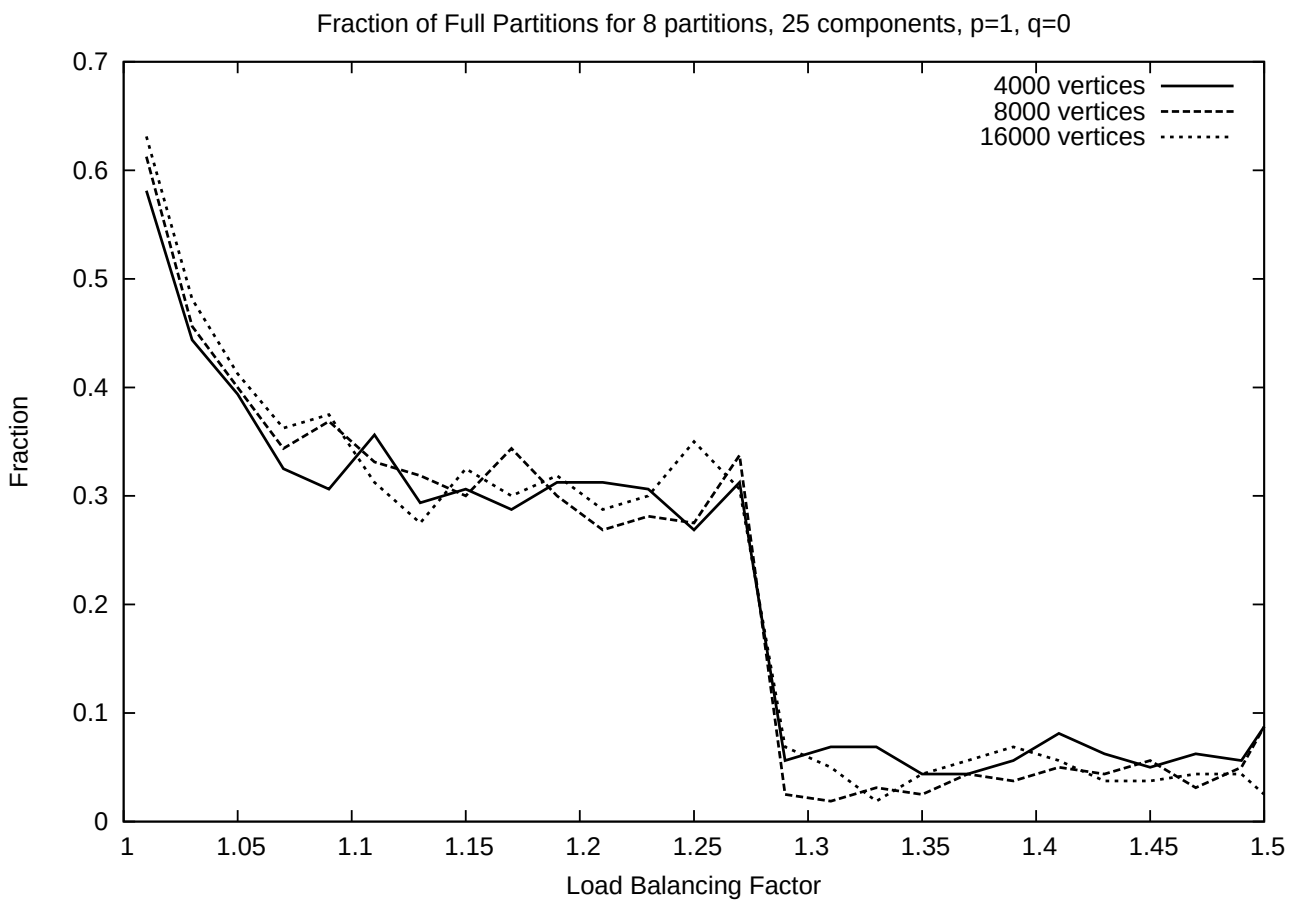

**Figure 1: Load balancing is not a function of the size of the graph**

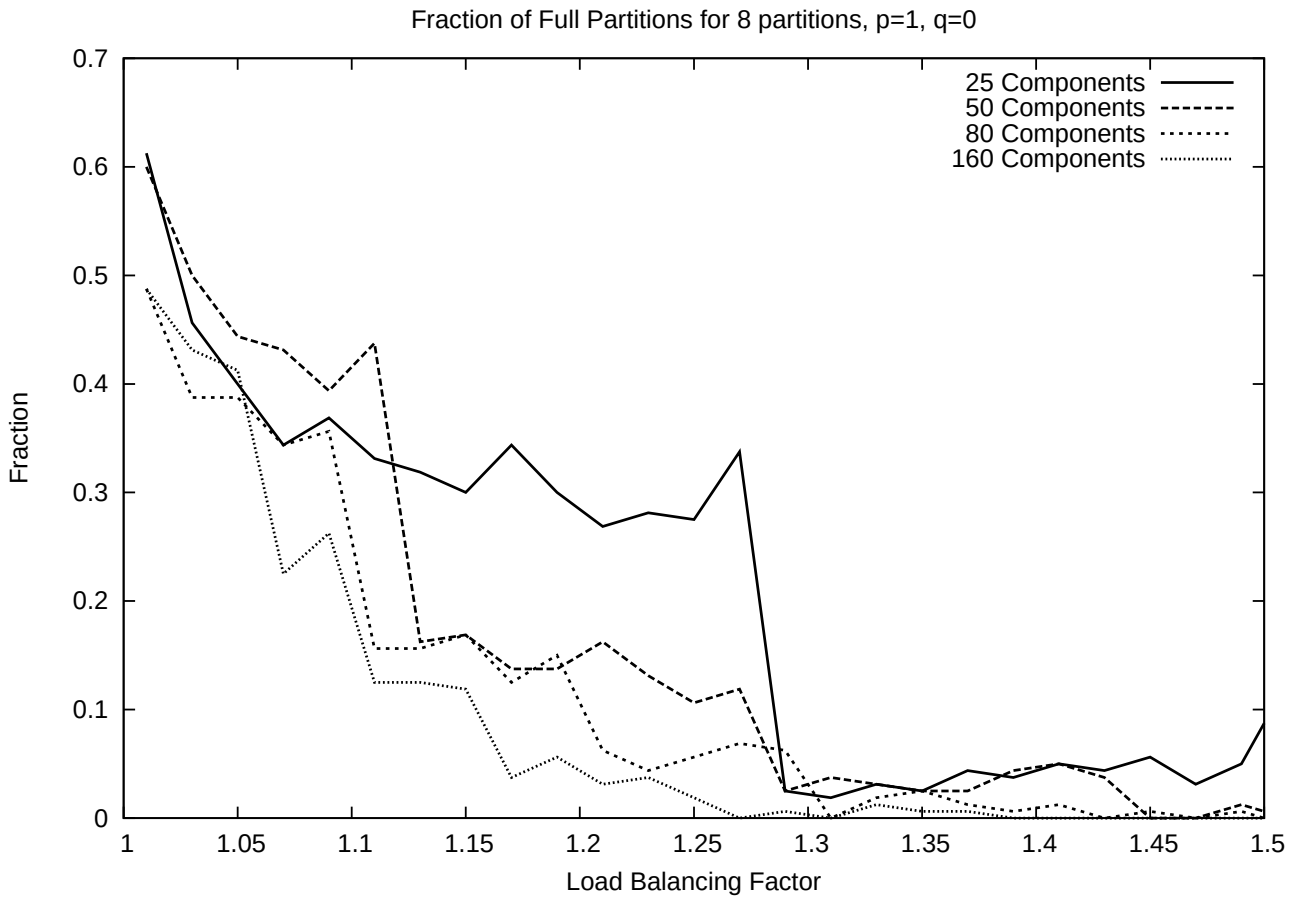

**Figure 2: Increasing the number of components improves the load balancing.**

## 6.1 Load Balancing Factor

Understanding the load balancing factor required is the first step to understanding the other constraints. This is because if the load balancing factor is set too low, we will see this in the error calculations. To understand the slack required, we explore two settings of $p$ and $q$, $p = 1$ and $q = 0$ or $q = \frac{p}{6kl}$ where $l$, the number of components is larger than $k \log k$. Now, for each size graph, we run the algorithm 20 times and record the number of partitions that hit their capacity constraints. We also vary $l$ to understand how its relationship with $k$ affects the required slack.

We include 3 figures to demonstrate the relationship.

The first, Figure 1 shows the fraction of full partitions when $\epsilon$ is allowed to range from 0.01 to 0.5 for graphs of size 4,000, 8,000 and 16,000. There is no difference between the threshold point in these graphs. The second, Figure 2 shows that fixing $p$, $q$ and $k$ but increasing $l$, the number of components, yields significantly better load balancing factors. The third Figure 6 (in the Appendix) shows that whether $q = 0$ or $q = 0.002 = p/6kl$, the load balancing appears the same.

## 6.2 Density Requirement

Lemma 6 requires that each component have edge density at least $p > \frac{2\log n}{|C_i|}$. To explore whether this is necessary, we can fix values for $q$, $k$ and $l$ and let $p$ range above and below $\frac{2\log n}{|C_i|}$. For each run, we measure the error from the perfect solution by looking at the Euclidean distance between the length-$l$ vector of the values of $\max_{j\in k} |C_i \cap P_j|/|C_i|$ and the all-ones vector.

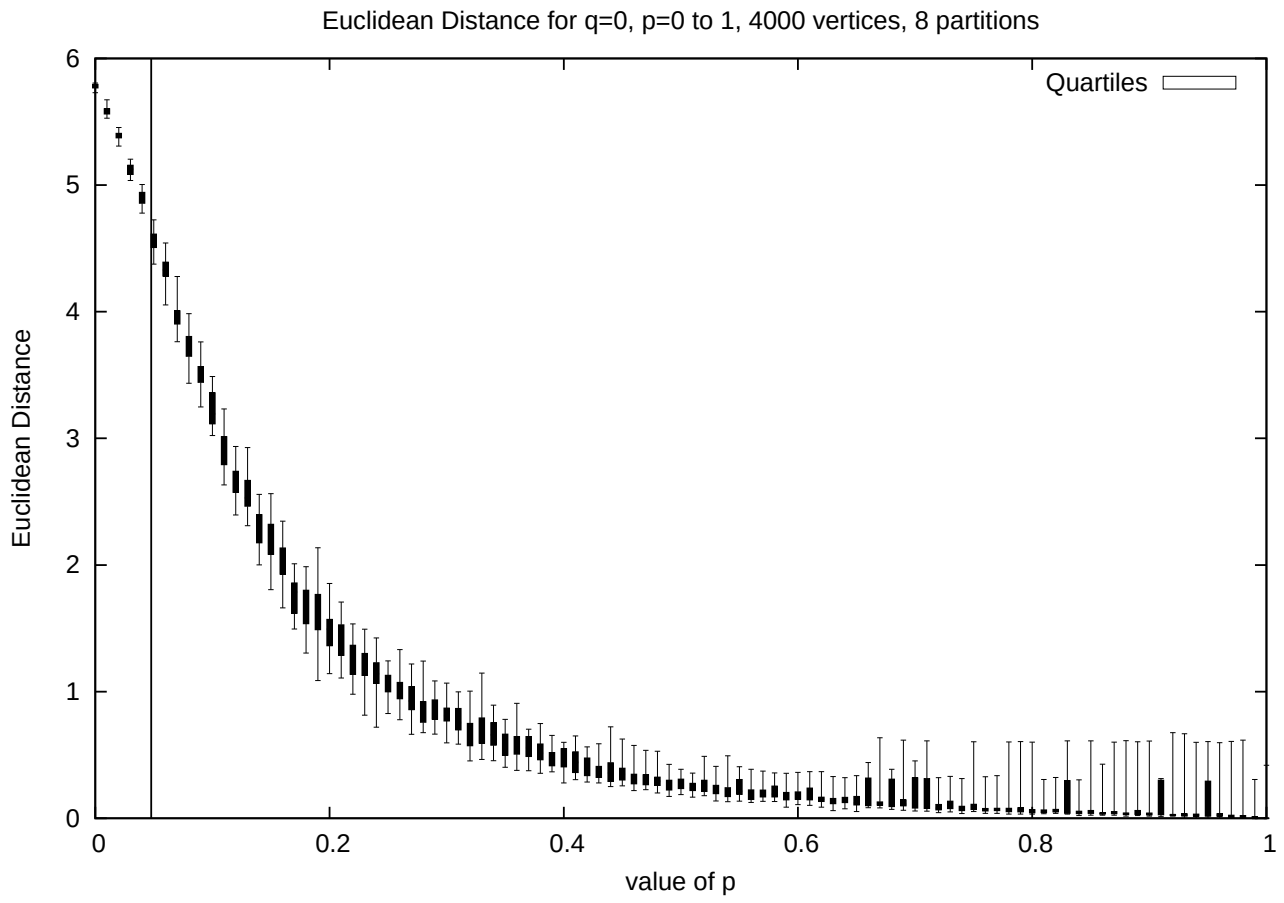

**Figure 3: For fixed $q, k, l$ values, as $p$ increases, the error in the partitioning generated drops to 0. The vertical bar marks the value required by the theorems.**

Though not pictured in Figure 3, the graph size shifts the 'elbow' of the graph to the left with a sharper transition, matching the bound of the theorem.

## 6.3 Constraints on $q$

As in the experiments to understand the density factor, we can also fix values for $p, k$ and $l$ and let $q$ range above and below $\frac{p}{6kl}$. Is the factor of $k$ necessary? We measure the error by Euclidean distance as above.

We clearly see the effect that increasing $q$ has on the algorithm's ability to recover the partitioning in Figure 4. While the value required by the theorems seems unnecessarily small (and can only be seen by zooming in on this page), dropping the required factor of $k$ and using $q = 0.02$ obtains an average error of only 0.07 over 25 runs when the maximum error is 7.

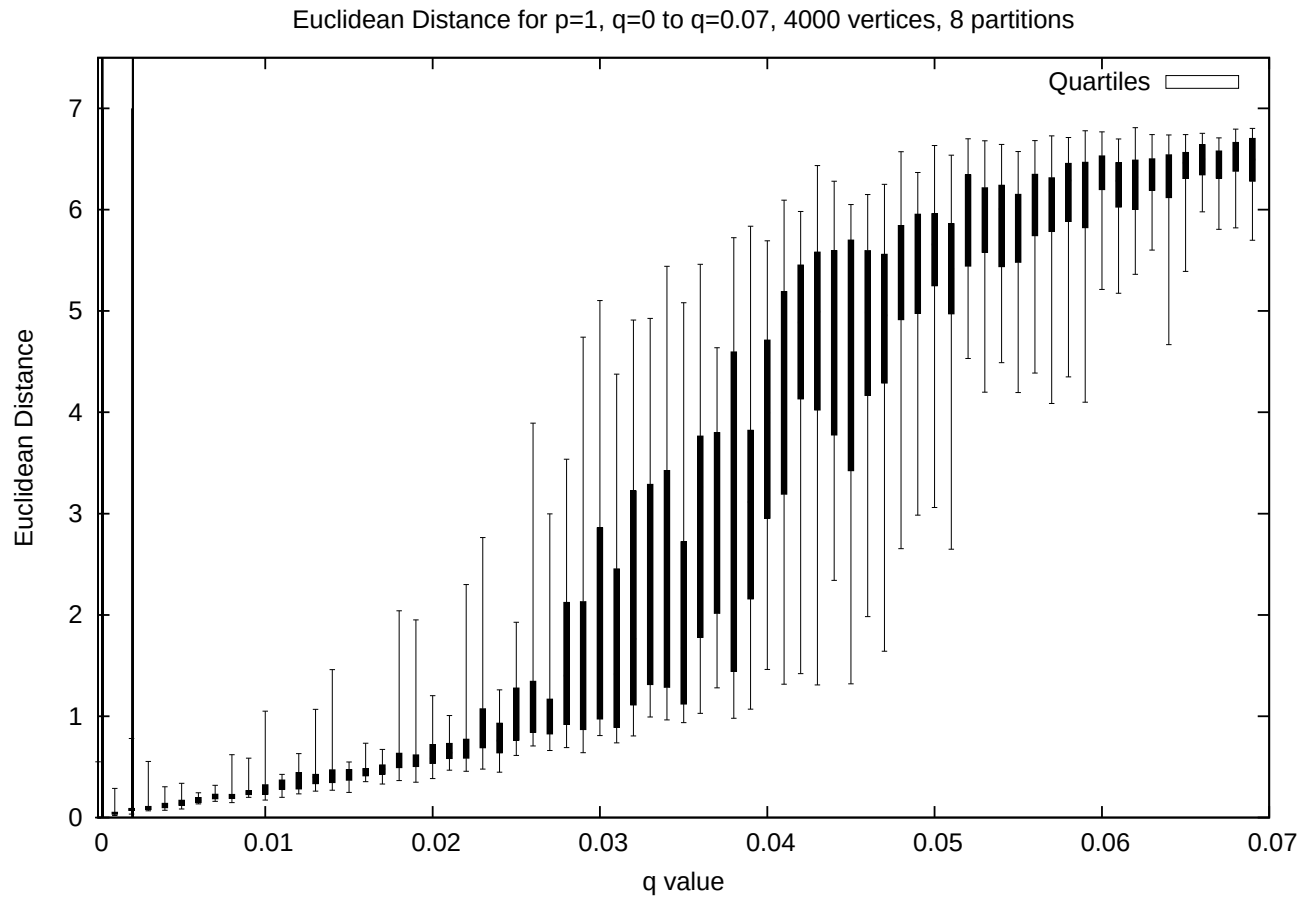

**Figure 4: For fixed $p, k, l$ values, as $q$ increases, the error in the partitioning increases from 0 to maximum error. The leftmost bar at 0.00026 marks the theorems' requirement, while the second at 0.0021 is $q = p/6l$.**

## 6.4 Convergence Rate

The values given by the Theorems in [11] about the rate of convergence imply a somewhat pessimistic bound - $q = O((k^{2.4}\log l)^{-1})$. We can evaluate this bound by fixing $p$, $q$, $k$ and $l$ and letting the size of the graph grow. As it grows, we can measure the Euclidean distance to find how quickly it is able to obtain good results in terms of recovering the partitioning.

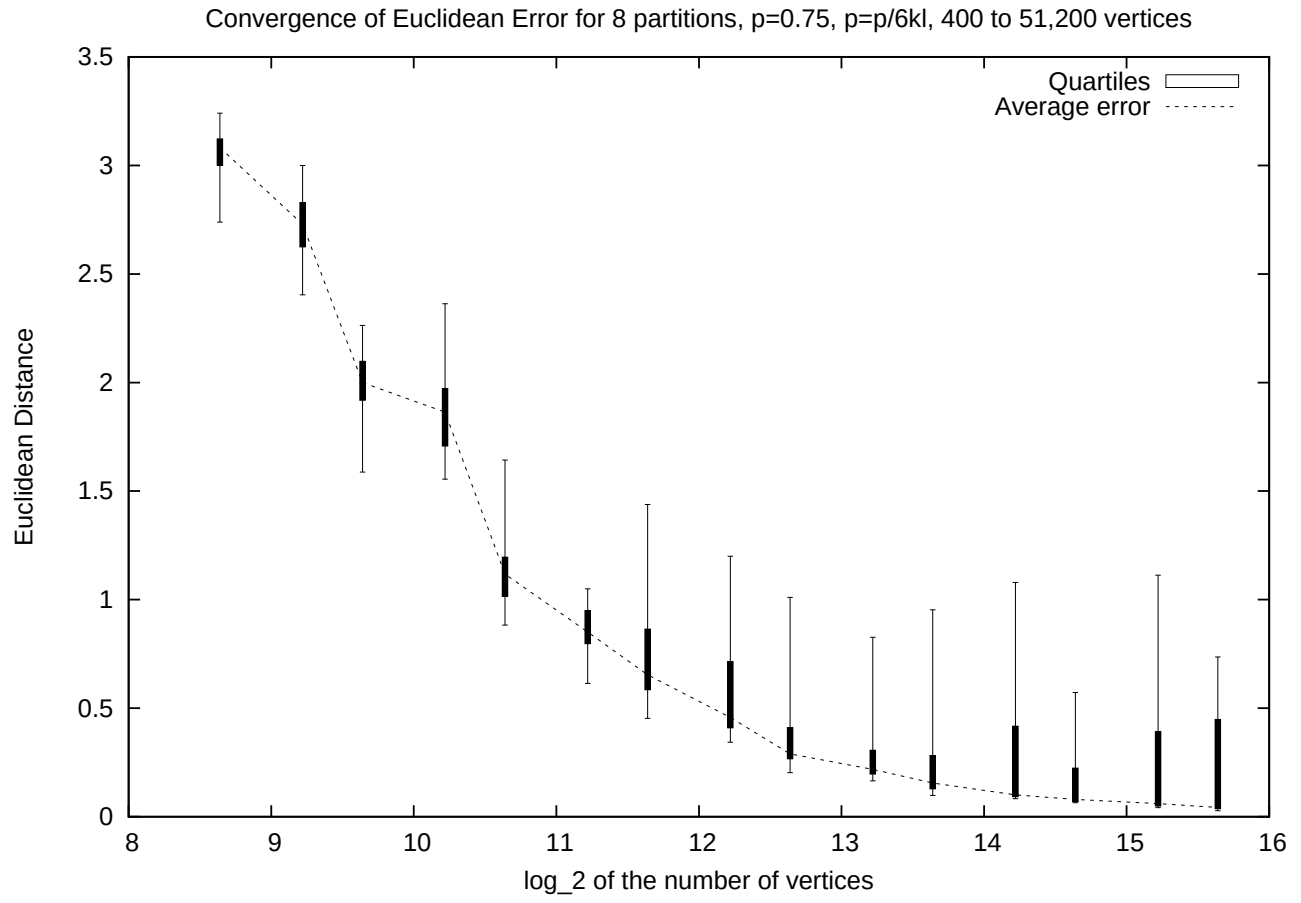

**Figure 5: This graph shows that for fixed $p, k, l, q$ values, as the size of the graph increases, the error in the partitioning generated drops to 0.**

The settings for the algorithm in Figure 5 were $p = 0.75, q = \frac{p}{6kl}, k = 8, l = 100$. The graph size range from 400 to 51,200 vertices. We see that as the size of the graph increases, the euclidean distance from the optimal partitioning solution quickly drops. For 51,200 vertices, the median error for 25 runs is only 0.04. This is despite the fact that the theorem required that $q < 0.000013$ whereas we used $q = 0.00015625$.

## 7. CONCLUSIONS AND FUTURE WORK

We have studied two simple greedy algorithms for streaming balanced graph partitioning. We first showed lower bounds on the possible approximation ratio obtainable by any algorithm and then analyzed two variants of a randomized greedy algorithm on a random graph model with embedded balanced $k$-cuts. On these graphs we were able to explain previous experimental results showing that the arg max Greedy algorithm is able to recover a good partitioning while the Proportional Greedy variant is not. Our proof connects the greedy algorithms with finite Polya urn processes and exploits concentration results about those processes.

There are several future directions. The first is to improve the parameters of the analysis. The experiments show that the algorithm continues to work with larger amounts of noise than that allowed by our theorems. The experiments in [25] show that the algorithm performs well on other random graph models like Watts-Strogatz. Explaining this and proving results about the approximation ratio is an interesting question.

Another direction is that in [25], additional stream orderings were studied. Experimentally, the algorithms tested all performed better on both of these orderings than the random ordering. An interesting open question is to develop techniques for analyzing streaming graph algorithms on BFS and DFS orders.

## APPENDIX

## A. HIGH PROBABILITY BOUNDS FOR LEMMA 6

The experiments justify the assumption that we only need the following two statements to hold with constant probability:

$$E_i^{(t)} > p|C_i|\frac{t}{n} - \sqrt{p|C_i|\frac{t}{n}}$$

$$B^t < ql|C_i|\frac{t}{n} + \sqrt{ql|C_i|\frac{t}{n}}$$

Requiring each to hold with probability $1-\delta$ increases the gap required from $p > 3(k+\sqrt{k}+1)kql$ by adding a dependency on $\delta$. In particular, redoing the calculations, we have that

$$p|C_i|\frac{t}{n} - \sqrt{\log(1/\delta)p|C_i|\frac{t}{n}} > k$$

exactly when

$$t > \frac{n}{p|C_i|}(k + \log(1/\delta)/2 + \sqrt{k\log(1/\delta) + (\log(1/\delta)^2/4})$$

Similarly,

$$ql|C_i|\frac{t}{n} + \sqrt{\log(1/\delta)ql|C_i|\frac{t}{n}} < 1$$

exactly when

$$t < \frac{n}{ql|C_i|}(1 + \log(1/\delta)/2 - \sqrt{\log(1/\delta) + (\log(1/\delta)^2/4})$$

Solving these two equations as in Lemma 7 gives us a similar relationship that $p > f(\delta)kql$.

## B. CALCULATION OF $Q$ FOR LEMMA 7

In order to prove Lemma 7 we need to understand for a given setting of $p$ and $q$ how much interaction between the components there is at the $t^{th}$ vertex. In particular, for the $t^{th}$ vertex, we expect that there will be $p\frac{t}{l}$ edges from that vertex to its own component (good edges) and $q\frac{(l-1)t}{l}$ edges to other components (bad edges). Provided $t < \frac{l}{q(l-1)}$, we do not expect any bad edges so the components do not interact at all.

When we do begin to see bad edges, we can appeal to Lemma 4. If it is the case that for the given component, one partition contains a $1/2+x$ fraction of the component that has arrived to this point, and all other partitions split the remaining $1/2-x$ fraction then we can argue that the bad edges do not affect the concentration of the process provided the arg max for the good edges is not changed by the addition of the bad edges. Specifically, we are concerned with $t = \frac{l}{q(l-1)} > \frac{1}{q}$ so we can find $x$ by solving:

$$(1/2+x)p\frac{t}{l} - \sqrt{(1/2+x)p\frac{t}{l}} > ((1/2-x)p\frac{t}{l} + \sqrt{(1/2-x)p\frac{t}{l}}$$

The above equation gives the distribution of the good edges at time $t$. Substituting that $t = \frac{1}{q}$ and there is only one bad edge, we need that

$$(1/2+x)\frac{p}{ql} - \sqrt{(1/2+x)\frac{p}{ql}} > ((1/2-x)\frac{p}{ql} + \sqrt{(1/2-x)\frac{p}{ql}}$$

This results in

$$x = \pm\sqrt{\frac{2(p/ql)^3 - (p/ql)^4}{4(p/ql)^4}}$$

From this, we can gather that a sufficient $\gamma$ value required for Lemma 4 is $\gamma = \frac{1}{2} - \sqrt{1/2(p/ql)}$. Lemma 4 gives a formula for translating this $\gamma$ into a $\delta$ value for Theorem 8. Solving for $\delta$ we get that

$$\delta = \frac{\gamma}{k-1-(k-2)\gamma}.$$

Plugging in our $\gamma$ value, we obtain that

$$\delta = \frac{1/2 - \sqrt{1/2(p/ql)}}{k-1-(k-2)(1/2-\sqrt{1/2(p/ql)})}.$$

We can simplify this by claiming that $\delta < \frac{1}{k}$ is sufficient.

The failure probability that we need to obtain from Theorem 8 for Lemma 4 is at most $\frac{c}{k^2 l}$ to use a union bound and still obtain a constant probability of success for the whole process. Therefore, we need to set $n_0' = n_0 + 2\log k + \log l$.

From here, we can obtain a number of balls thrown before we can obtain this level of concentration. In particular, we need $2^{x+z}n_0$ balls, where $x = \log_{1+\frac{\lambda-1}{5+4(\lambda-1)}} \frac{0.4}{\epsilon_0}$ and $z = \log_{\frac{2\lambda}{\lambda+1}} \frac{0.1}{\delta}$. The $x$ term allows us to obtain up to all-but-0.1 dominance, while the second improves the result to all-but-$\delta$ dominance. Therefore, if $k \leq 10$, then we only need $n_0 2^x$ balls. More generally, substituting that $\epsilon_0 = 1/5\lambda$ and $\delta = \frac{1}{k}$, this value becomes:

$$(2\lambda)^{1/\log_2(5\lambda/(1+4\lambda))}(0.1k)^{1/\log_2(2\lambda/(\lambda+1))}n_0'$$

The interesting thing about the process is that as more vertices arrive, the $\lambda$ value increases. From this, we can immediately claim that this equation dramatically over-estimates the number of vertices needed before 2 bins would obtain a state with all-but-$\frac{1}{k}$ dominance. In particular, for the $p$ and $q$ values required by Lemma 6, we have $p = 6klq$ so $\lambda$ reaches a value of $3k$ before we expect to see bad edges. Unfortunately, the best we can assume is that $\lambda = 2$ obtaining the following value:

$$4^6.578(0.1k)^2.4n_0' \approx 9127n_0'(0.1k)^{2.4}$$

It is certainly possible to set $q$ to $1/9127n'_0(0.1k)^{2.4}$ but it is a significantly different bound from $p > 6klq$.

## C. EXPERIMENTAL RESULTS

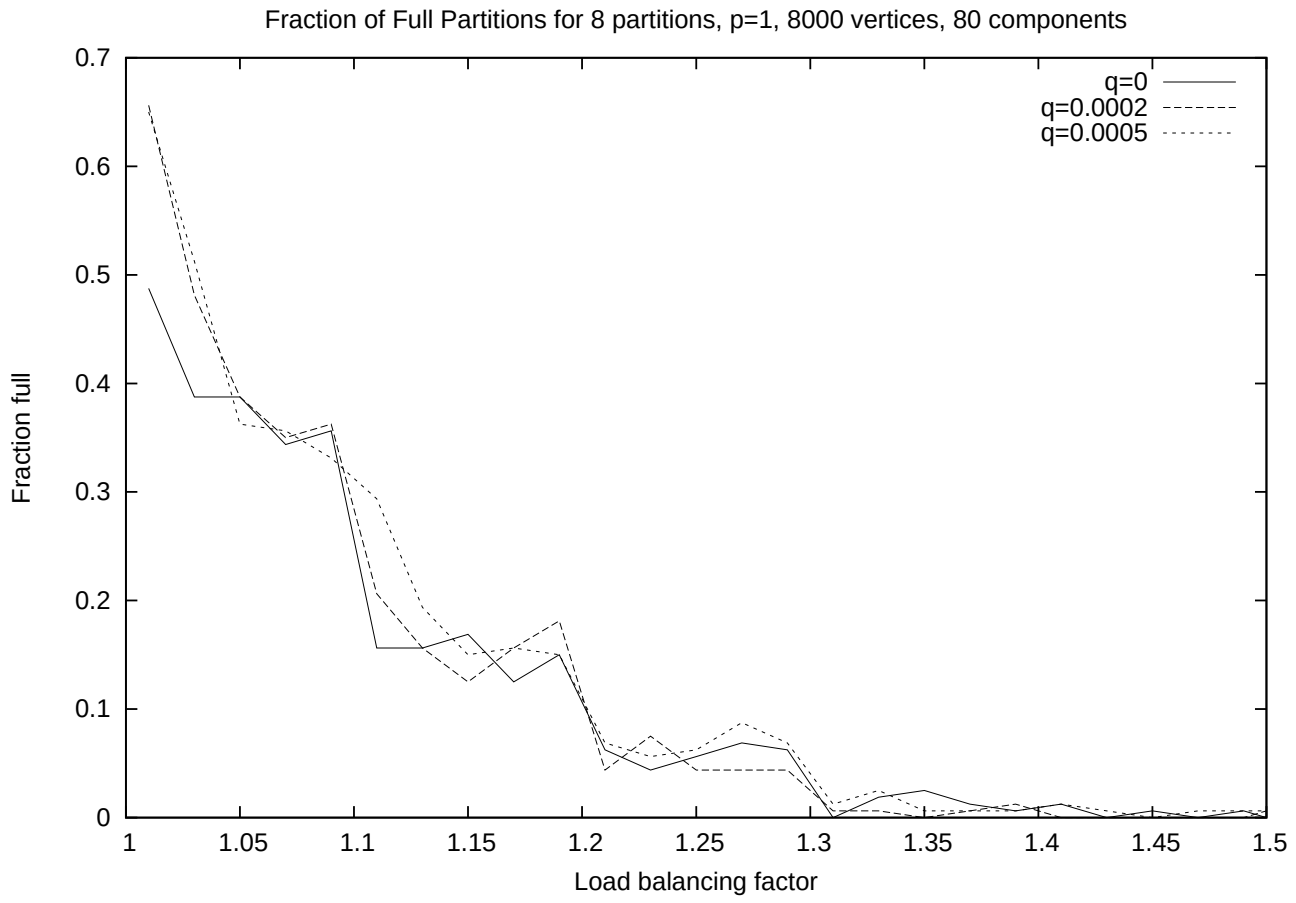

**Figure 6: $q$ does not play a large role in load balancing. Note that $q = 0.0005$ is above the threshold required by the theorems.**